\documentclass[onecolumn,epsfig]{elsart}
\usepackage{graphics} 
\usepackage{graphicx} % Include Fig. files
\usepackage{dcolumn} % Align table columns on decimal point
\usepackage{amssymb}

\newcommand{\AmS}{{\protect\the\textfont2A\kern-.1667em\lower.5ex\hbox{M}\kern-.125emS}}
\newcommand{\pt}{$p_{T}$ }

\def\Journal#1#2#3#4{{#1} {\bf #2}, #3 (#4)}
\def\NIMA{{ Nucl. Instrum. Methods} A}

\def\PLB{{ Phys. Lett.}  B}
\def\PRL{ Phys. Rev. Lett.}
\def\PR{ Phys. Rev.}
\def\PRC{{ Phys. Rev.} C}
\def\PRD{{ Phys. Rev.} D}
\def\ZPC{{ Z. Phys.} C}
\def\JPG{{ J. Phys.} G}

\def\bea{\begin{eqnarray}}
\def\eea{\end{eqnarray}}

\begin{document}
%\preprint{Intended for PLB}

\begin{frontmatter}

\title{Yields and elliptic flow of $d(\overline{d})$ and $^{3}He(\overline{^{3}He})$ in Au+Au collisions at $\sqrt{s_{_{NN}}} =$ 200 GeV}

\author[uic]{B.~I.~Abelev},
\author[pu]{M.~M.~Aggarwal},
\author[vecc]{Z.~Ahammed},
\author[jinr]{A.~V.~Alakhverdyants},
\author[kent]{B.~D.~Anderson},
\author[dubna]{D.~Arkhipkin},
\author[jinr]{G.~S.~Averichev},
\author[mit]{J.~Balewski},
\author[uic]{O.~Barannikova},
\author[uuk]{L.~S.~Barnby},
% ...

(STAR Collaboration)

% ...
\address[uuk]{University of Birmingham, Birmingham, United Kingdom}
% ...
\address[uic]{University of Illinois at Chicago, Chicago, Illinois 60607, USA}
% ...
\address[jinr]{Laboratory for High Energy (JINR), Dubna, Russia}
\address[dubna]{Particle Physics Laboratory (JINR), Dubna, Russia}
% ...
\address[kent]{Kent State University, Kent, Ohio 44242, USA}
% ...
\address[mit]{Massachusetts Institute of Technology, Cambridge, MA 02139-4307, USA}
% ...
\address[pu]{Panjab University, Chandigarh 160014, India}
% ...
\address[vecc]{Variable Energy Cyclotron Centre, Kolkata 700064, India}
% ...

%\address{}

\date{\today}% It is always \today, today,

\begin{abstract}
We present the $p_{T}$ spectra, elliptic flow ($v_2$) and coalescence
parameters $B_{2}$ for $d$, $\overline{d}$ ($1<p_{T}<4$ GeV/c) and
$B_{3}$ for $^{3}He$, $\overline{^{3}He}$ ($2<p_{T}<6$ GeV/c) produced
at mid-rapidity in Au+Au collisions at $\sqrt{s_{_{NN}}} = 200$
GeV. The results are measured in the STAR experiment at RHIC. The
spectra of the light nuclei show softer $p_T$ distributions than
calculations from a blast-wave model in which the parameters were
fixed from pion, kaon and proton $v_2$($p_T$) and $p_T$
distributions. The coalescence volume is found to track with pion HBT
results for different collision geometries. The $v_2$ measurement for
$d(\overline{d})$ as a function of transverse momentum $p_T$ is found
to follow an approximate atomic mass number ($A$) scaling while that
of $^{3}He(\overline{^{3}He})$ deviates slightly from the scaling. A
negative $v_{2}$ has been observed for $\overline{d}$ at low $p_{T}$,
consistent with large radial flow in Au+Au collisions.

\end{abstract}

\end{frontmatter}
%\maketitle

\section{Introduction}
In relativistic heavy ion collisions, light nuclei and anti-nuclei are formed through coalescence of produced nucleons and anti-nucleons or participant nucleons~\cite{final_state, coalescence1,coalescence2}. Since the binding energy is small, this formation process can only happen at a late stage of the evolution of the system when interactions between nucleons and other particles are weak. This process is called final-state coalescence ~\cite{final_state, BA_ref}. The coalescence probability is related to the local nucleon density. Therefore, the production of light nuclei provides a tool to measure collective motion and freeze-out properties, such as particle density~\cite{xuliu} and correlation volume.

Invariant yields for the production of nuclei can be related~\cite{final_state} to the primordial yields of nucleons by Equation~\ref{BA}.
\begin{equation}\label{BA}
    E_{A}{{d^{3}N_{A}}\over{d^{3}p_{A}}} =
B_{A}(E_{p}{{d^{3}N_{p}}\over{d^{3}p_{p}}})^{Z}
(E_{n}{{d^{3}N_{n}}\over{d^{3}p_{n}}})^{A-Z} \approx
B_{A}(E_{p}{{d^{3}N_{p}}\over{d^{3}p_{p}}})^{A}
\end{equation}
where $N_A$, $N_p$, and $N_n$ denote the yields of the particular nucleus, and of its constituent protons and neutrons, respectively. $B_{A}$ is the coalescence parameter. $E{{d^{3}N}\over{d^{3}p}}$ is the invariant yield of nucleons or nuclei; $A$ and $Z$ are the atomic mass number and atomic number, respectively; $p_{A}$, $p_{p}$ and $p_{n}$ are the momenta of the nuclei, protons and neutrons, respectively, where $p_{A}= A\cdot p_{p}$ is assumed. The coalescence parameter, $B_{A}$, is related to the freeze-out correlation volume~\cite{final_state}: $B_{A}\propto V_f^{1-A}$. For an expanding fireball, the effective homogeneous coalescence volume decreases with transverse mass and temperature ($M_{t}/T$) as detailed in Ref.~\cite{coalescence1}. 

On the other hand, a blast-wave model is often used to describe the spectra of identified particles produced in relativistic heavy ion collisions~\cite{BW,STAR_proton_spectra}. In this model, the particle spectra are determined by global parameters (temperature, flow profile) and particle mass. Up to now, hydrodynamic model simulations or blast-wave fits to the data only included elementary particles without any composite elements, such as nuclei~\cite{BW}. Experimental measurements can be used to provide insights into whether the homogeneous volume (as in a coalescence picture) or mass (as in a blast-wave model) has a bigger influence on the yields of nuclei in heavy ion collisions.

In the case of deuteron production at low \pt, $B_2\propto 1/V_f$. The ratio of deuteron yield over proton yield at the same $p_T/A$ (denoted as $d/p$, and $\bar{d}/\bar{p}$ for anti-deuteron), is proportional to the baryon density~\cite{xuliu,fqwang}. This is analogous to the deuteron to hydrogen ratio ($D/H$) measurements of Big-Bang nucleosynthesis (BBN), which is a very sensitive probe of baryon abundance in the early universe~\cite{BBN1,PDGBBN}. Although both processes are sensitive to the baryon density, the processes themselves are very different. In coalescence, a proton and a neutron form a deuteron due to the overlap of the quantum wave function in a dilute QCD medium, while in BBN the deuteron production is through p(n,$\gamma$)D photo-production. Nevertheless, it is interesting to investigate whether the coalescence at much higher temperature and density, which existed also at the pre-BBN stage of the early universe, produces a dramatically different deuteron abundance when compared to the photo-production of deuterium in the standard BBN model.

Coalescence has been generalized as a mechanism for partons to form hadrons at an early stage of heavy ion collisions~\cite{quarkcoalescence,coal_dyn}. It is experimentally difficult to study how local correlations and energy/entropy play a role in coalescence at the partonic level since the constituents are not directly observable. By studying the spectra and elliptic flow of nuclei and comparing to those of their constituents (nucleons), we have a better handle on how coalescence happens. It has been found that elliptic flow ($v_{2}$) of hadrons is very sensitive to their underlying partonic degrees of freedom~\cite{generalv2,quarkcoalescence}. An analog in nucleus
production is to test if nucleus elliptic flow scales with the atomic mass number ($A$). Deviations may point to a non-uniform nucleon distribution with respect to the collision geometry or large nuclear density fluctuations. This will provide valuable data for studying the freeze-out dynamics and coalescence mechanism in general.

Previously, measurements of nucleus production at RHIC suffered from low statistics. The first observation of anti-helium ~\cite{starnucleus} at RHIC and measurements of coalescence properties ~\cite{phenixnucleus} with (anti-)deuterons at RHIC have been previously published. More recently, increasing RHIC luminosity and detector upgrades allow more precise measurements of yield and elliptic flow of nuclei. In this paper, we present spectra and $v_{2}$ of the deuteron, $^{3}He$, and their anti-particles ($\overline{d}$ and $\overline{{^{3}He}}$). We compare volumes extracted from nucleus
production and from pion HBT. First measurements of $^{3}He$ $v_{2}$ and high statistics $\overline{d}$ $v_{2}$ at low $p_{T}$ will be presented and their implications will be discussed.

\section{Experiment and analysis}

\begin{figure}
\begin{center}
\includegraphics[scale=0.7]{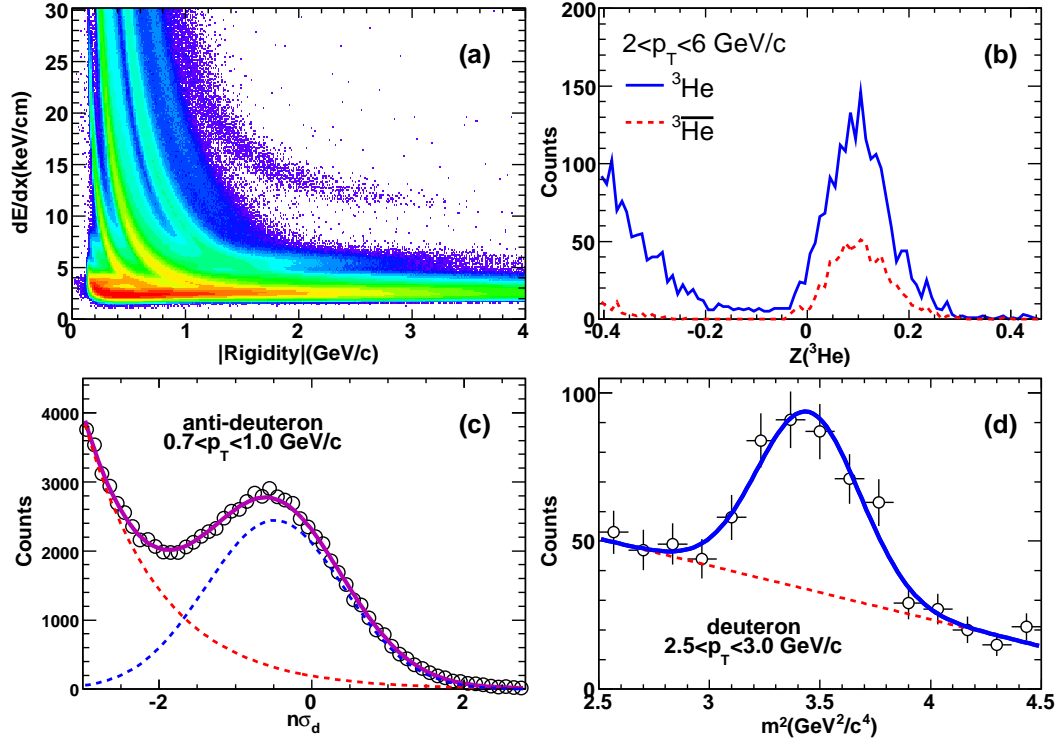}
\caption{(a) TPC $dE/dx$ as a function of $|$rigidity$|$.  (b) $Z$ ($Z=\log((dE/dx)|_{measure}/(dE/dx)|_{predict})$) distribution of $^{3}He$ (solid line) and $\overline{^{3}He}$ (dashed line). (c) $n\sigma_{d}$ distribution of $\overline{d}$ at $0.7<p_{T}<1.0$ GeV/c with a Gaussian fit plus an exponential background. (d) The
distribution of squared mass ($m^{2}=(p/\beta/\gamma)^{2}$) for $d$ from the TOF after TPC $dE/dx$ selections at $2.5<p_{T}<3.0$ GeV/c, with a Gaussian fit plus a linear background.}
\label{Tech_figure}
\end{center}
\end{figure}

The data presented here are obtained from the Time Projection Chamber (TPC)~\cite{STAR} and the Time-Of-Flight (TOF) detectors~\cite{JWu} in the STAR experiment at RHIC in the year 2004. The TOF detector system was a prototype module covering $\pi/30$ rad in azimuth and $-1.0<\eta <0$ in pseudorapidity. These analyses used a  data sample of 25 million central triggered events (0-12\% centrality) and 24 million minimum-bias triggered events (0-80\% centrality) and TOF information is available for 16 million central and 15 million minimum-bias events. The selection of centrality is based on the Glauber model and is described in Ref.~\cite{centrality}. Measurements of the ionization energy loss ($dE/dx$) of charged tracks in the TPC gas are used to identify protons, deuterons, ${}^{3}He$ and their anti-particles. By combining the particle identification capability of $dE/dx$ from the TPC and velocity from the TOF, pions and protons can be identified in $0.3<p_T<12$ GeV/c ~\cite{Ming,STAR_proton_spectra}. In our analysis, deuterons and anti-deuterons are identified by TPC for $p_{T}<1$ GeV/c and by TOF in the range $1<p_{T}<4$ GeV/c. However, at low $p_T$ ($<1$ GeV/c), primary deuterons are overwhelmed by background from knock-out deuterons from the beam pipe and inner detector material, which are difficult to separate from collision products. As a result, only anti-deuterons are counted as collision products.  Because the absorption effect and detector efficiency for $\overline{d}$ at $p_T<1$ GeV/c are not well simulated in GEANT, and other produced particles in the same event produce large deuteron background from detector material and beam pipe, only identified $\overline{d}$ in this $p_T$ range are used for elliptic flow calculation. The TPC is used to identify ${}^{3}He$ and $\overline{{}^{3}He}$ in the range $2<p_{T}<6$ GeV/c.

Figure~\ref{Tech_figure} illustrates the particle identification techniques and methods. Panel (a) shows the $dE/dx$ of charged tracks as a function of $|$rigidity$|$ ($rigidity=momentum/charge$) measured by the TPC at $-1<\eta<1$. Panel (b) shows $Z$ ($Z=\log((dE/dx)|_{measure}/(dE/dx)|_{predict})$) distribution for $^{3}He$ and $\overline{^{3}He}$ signals, where $(dE/dx)|_{predict}$ is a function accounting for the curvature of $dE/dx$ versus momentum \cite{Ming, bichsel}. With tight track quality selections imposed by requiring the number of TPC points in a track to be greater than 25 and the distance of closest approach to the event vertex to be less than 1 cm, the $^{3}He(\overline{^{3}He})$ signals are essentially background free. Panel (c) shows $n\sigma_{d}$ (standard Gaussian deviation from $(dE/dx)|_{predict}$) distribution for $\overline{d}$ at $0.7<p_{T}<1.0$ GeV/c. The signal was fit with a Gaussian function and an exponential background. As can be seen in panels (b) and (c) of Figure 1, the peak centroids in $Z$ and $n\sigma_d$ are shifted from zero. The large shift in $^3He$ is related to the higher ionization density for particles with charge number $>$1. See Ref.~\cite{Yichun} and Fig. 14 of Ref.~\cite{bichsel}. It may be changed by the effects during ion collection. Since this offset does not affect the extracted yields, no attempt was made to more carefully match the overall magnitude of $dE/dx$. Panel (d) shows $m^{2}$ distribution for $d$ at $2.5<p_{T}<3.0$ GeV/c measured by the TOF after the $dE/dx$ cut ($|n\sigma_{d}|<2$, in which $n\sigma_{d}$ denotes the standard Gaussian $dE/dx$ deviation from the expected deuteron value.)~\cite{STAR_proton_spectra,Ming}. The signal was fit with a Gaussian function plus a linear background. The acceptance and tracking efficiencies were studied by Monte Carlo GEANT simulations of the STAR detector~\cite{STAR}.
%We note that in an ideal situation, the $n\sigma_{^{3}He}$ (standard Gaussian deviation $dE/dx$ from the expected $^{3}He$ value) should peak at zero with a width of unity. The deviation from zero seen in the peak in panel (b) shows that the Bichsel function~\cite{Ming, bichsel} does not match the calibrated $dE/dx$ precisely. However, this should not affect the yields because they are counted by integrating the distributions over the background-free helium mass peak. 

\section{Results}

\subsection{Spectra and the coalescence parameters}
\begin{figure}
\begin{center}
\includegraphics[scale=0.7]{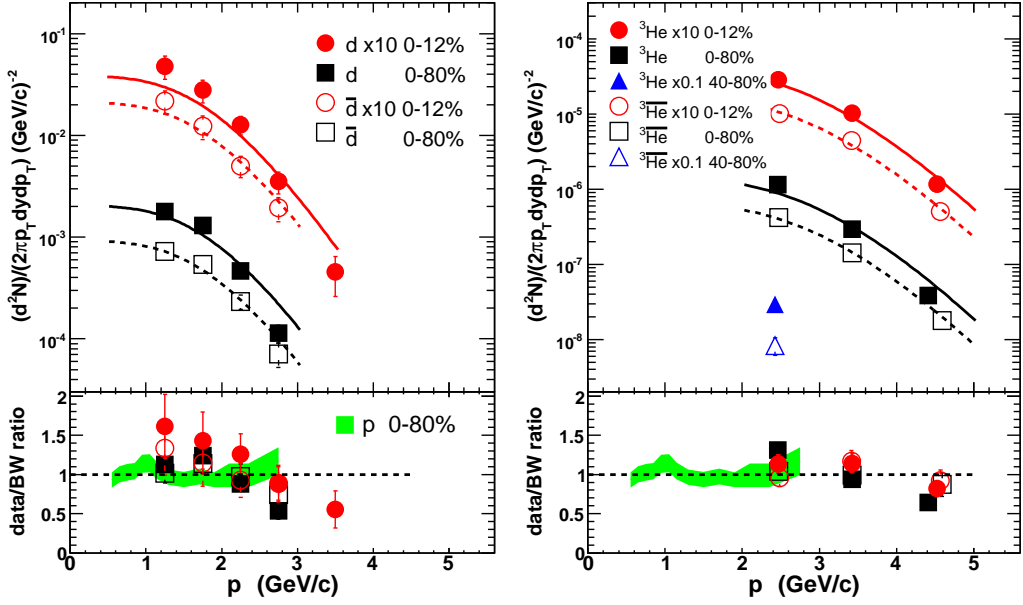}
\caption{The left (right) plot shows $d$ and $\overline{d}$ ($^{3}He$ and $\overline{^{3}He})$ spectra, with a comparison to the blast-wave model calculations.  In each plot, the upper panel shows the $p_{T}$ spectra, with the solid symbols and open symbols representing the particles and anti-particles, respectively. The corresponding blast-wave calculations are shown by solid and dashed lines. The lower panel in each plot shows the data divided by the blast-wave calculation. The bands show the same ratio for protons. Errors are statistical only.}
\label{spectra_B2B3_figure}
\end{center}
\end{figure}

The blast-wave (BW) model has been successfully applied to reproduce the $\pi$, $K$, $p$, and $\Lambda$ spectra~\cite{STAR_proton_spectra}, and elliptic flow as well as $\pi$ HBT correlations as a function of transverse momentum and centrality~\cite{BW}. This model calculates particle production properties by assuming a parameterized fireball expansion after the collision. The model also assumes local thermal equilibrium with an expansion velocity profile as a function of transverse radius, modulated by an azimuthal density distribution~\cite{BW}. Predictions from this model for nucleus yields and elliptic flow were compared to our data. Figure~\ref{spectra_B2B3_figure} shows the $p_{T}$ spectra of $d(\overline{d})$ and $^{3}He(\overline{^{3}He})$. The parameters obtained from using the blast-wave model~\cite{BW} to describe the spectra and $v_{2}$ of $\pi$, $K$, $p$ are used to calculate $d(\overline{d})$ and $^{3}He(\overline{^{3}He})$ spectra, with the results shown in Figure~\ref{spectra_B2B3_figure}. The spectra and $v_{2}$ of $\pi$, $K$, $p$ are described simultaneously by a single set of parameters, which is shown in Table \ref{BWpar} (only the 0-80\% centrality parameters are shown in the table). In the BW model, the nuclei are simply treated as heavier particles emitted from the fireball. The parameters used in the BW description are $T$, $\rho_{0}$, $\rho_{2}$, $R_{x}/R_{y}$, $\tau_{0}$ and $\Delta\tau$, where $T$ is the temperature. The freeze-out distribution is assumed to be infinite along the beam direction ($z$ direction) and elliptical in the transverse direction ($x$-$y$ plane), $R_{x}$ and $R_{y}$ giving the radii of the ellipse. The parameters $\rho_{0}$ and $\rho_{2}$ are the zeroth and second order factor of the flow boost along the direction perpendicular to the transverse ellipse, respectively. The source is assumed to emit particles over a finite duration in longitudinal freeze-out proper time ($\tau=\sqrt{t^{2}-z^{2}}$) peaked at $\tau_{0}$ with a Gaussian distribution width $\Delta\tau$.  A more detailed description and definition of the parameters and the BW formulae are given in Ref ~\cite{BW}.

In general, the spectra are similar to the model calculations. The ratio between data and model is shown in the bottom panels of Figure~\ref{spectra_B2B3_figure}. The model calculation tends to give harder spectra than seen in the data. This indicates that the coalescence process results in nuclei $p_{T}$ spectra which do not follow the simple mass dependence expected in the BW model, implying a deviation from simple thermal production.

\begin{center}
\begin{table}[h!b!p!]
\caption{BW parameters: see text for details.}
%\begin{tabular*}{0.9\textwidth}{|c|c|c|c|c|c|}
\begin{tabular}{|c|c|c|c|c|c|}
  \hline
  $T$(MeV) & $\rho_{0}$ & $\rho_{2}$ & $R_{x}/R_{y}$ 
   & $\tau_{0}$(fm/c) & $\Delta\tau$(fm/c)  \\
  \hline
  124.2  &
  0.88 &
  0.061 &
  0.89 &
  9.2  & 0.03  \\
  %$\pm$ 1.9 &
  %$\pm$ 0.01 &
  %$\pm$ 0.002 &
  %$\pm$ 0.003 &
  %(%fixed) & (fixed) \\
   \hline
\end{tabular}
\label{BWpar}
\end{table}
\end{center}

\begin{figure}
\begin{center}
\includegraphics[width=0.84\textwidth,keepaspectratio]{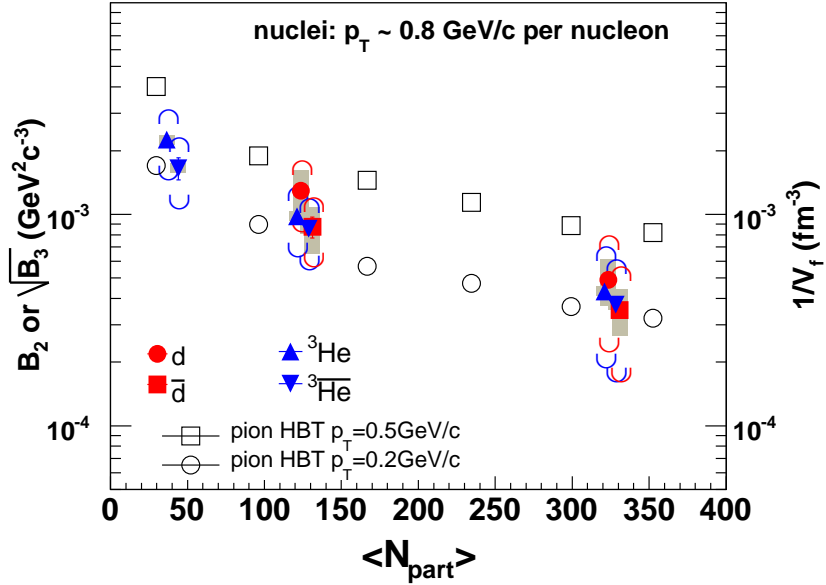}
\caption{$B_2$ and $\sqrt{B_3}$ together with $\pi^{\pm}$ HBT volume as a function of collision centrality ($\langle N_{part}\rangle$) in Au+Au collisions. HBT volume is calculated from the HBT correlation lengths along the longitudinal and transverse directions. The gray bands represent systematic errors and the brackets show the uncertainties from the feed-down estimation.}
\label{HBT_B2B3}
\end{center}
\end{figure}

Figure~\ref{HBT_B2B3} shows the coalescence parameters $B_{2}$ and $\sqrt{B_{3}}$, which are derived from $p$($\overline{p}$), $d(\overline{d})$ and $^{3}He$($\overline{^{3}He}$) spectra by Equation~\ref{BA}. Here the proton and anti-proton spectra are taken from Ref. ~\cite{STAR_proton_spectra}. The $p(\overline{p})$ spectra have been corrected for feed-down from $\Lambda(\overline{\Lambda}$) and $\Sigma^{\pm}$ weak decays ~\cite{STAR_proton_spectra}. As
mentioned before, $B_{A}\propto V_f^{1-A}$, so $B_{2}\propto1/V_f$ and $B_{3}\propto 1/V_f^2$. Therefore $B_{2}$ for $d$($\overline{d}$) should be proportional to $\sqrt{B_{3}}$ for $^{3}He$($\overline{^{3}He}$) if the correlation volumes for $d$($\overline{d}$) and $^{3}He$($\overline{^{3}He}$) are similar. Both $B_{2}$ and $B_{3}$ show strong centrality
dependence. In more central collisions, the smaller coalescence parameter indicates that the correlation volume at thermal freeze-out is larger than for peripheral collisions. This also means that the correlation length in nucleus coalescence grows with the system size.
 
Figure~\ref{HBT_B2B3} also shows a comparison with the results of pion HBT data. To calculate the freeze-out volume from HBT measurements, we use the following expression: $$V_f = (2\pi)^{3/2}\times R_{long}\times R^{2}_{side}$$ where $V_f$ is the freeze-out volume~\cite{coalescence1} and $R_{long}$ and $R_{side}$ are the longitudinal and sideward radii, respectively, assuming a density distribution of Gaussian shape in all three dimensions. The $R_{long}$ and $R_{side}$ values are taken from Ref.~\cite{BW,pbar_flow} ($k_T = 0.2$ GeV/c). The $d$($\overline{d}$) and $^{3}He$($\overline{^{3}He}$) transverse momentum ranges are $1.5<p_T<2.0$ GeV/c and $2.0<p_T<2.5$ GeV/c, respectively. The HBT data is chosen at the closest $p_T$ to the $p_T/A$ for the nuclei coalescence data throughout all of the centrality bins. The observations that the $B_2$ and $\sqrt{B_3}$ coalescence parameters are proportional to $1/V_f$ from pion HBT over the full range of centrality considered indicates that the freeze-out volume for the nuclei is closely related to that for pions.

In the coalescence model~\cite{coalescence1}, the proportionality factors quantitatively connecting the $B_{2}$ and $B_{3}$ parameters to the homogeneous volume depend on flow profile, temperature, correction factors due to quantum wave functions, and other detailed assumptions of the coalescence models. A precise extraction of these model-dependent factors from data will be possible in the future when the large uncertainty on $B_{2}$ and $B_{3}$ is reduced with the improvement of weak-decay feed-down correction to the proton yields~\cite{STAR_proton_spectra}.

\subsection{Elliptic flow parameter $v_{2}$}
The elliptic flow parameter, $v_{2}$, is the second order Fourier coefficient of the azimuthal distribution of the produced nuclei relative to the reaction plane of the initial nucleus-nucleus collision. The event-plane method was used to obtain the $v_2$ of nuclei~\cite{v2method}, with the
event plane resolution used for correction (calculated using the sub-event method~\cite{v2method}) of 76\% for minimum bias triggered events and 68\% for central triggered events. In the following discussions only statistical errors will be shown since the low statistics for deuteron and helium nuclei result in these errors being much larger than the estimated systematic errors.

\begin{figure}
%\begin{center}
\includegraphics[scale=0.6]{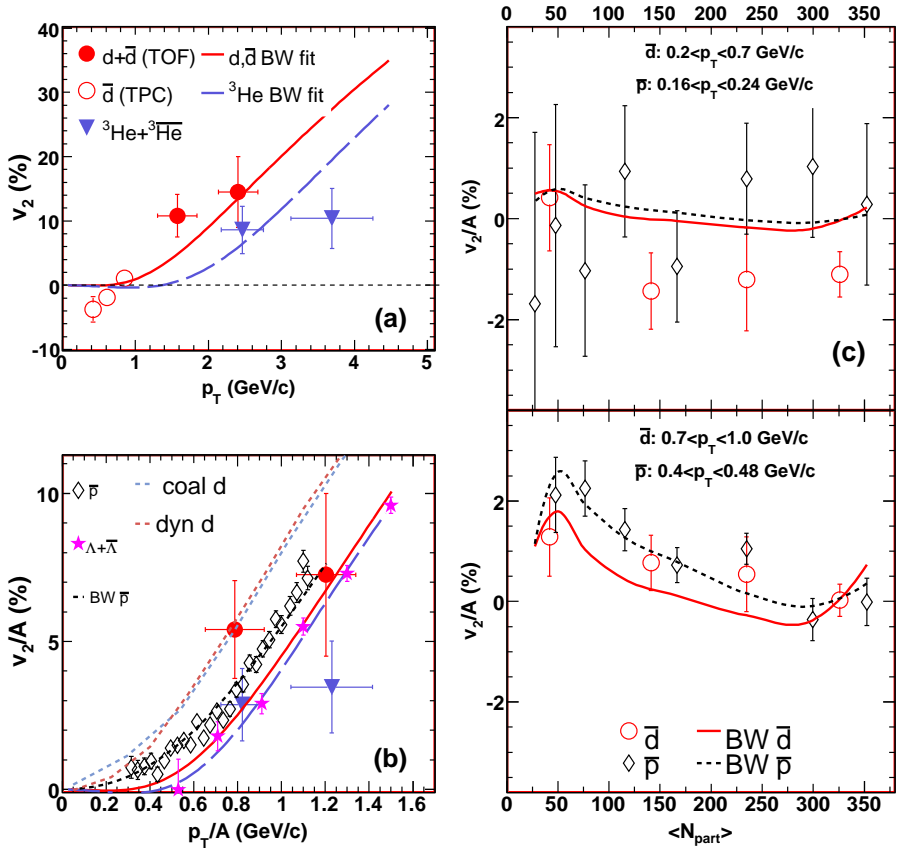}
\caption{(a) The elliptic flow parameter $v_{2}$ from minimum bias collisions as a function of $p_{T}$ for $^{3}He+\overline{^{3}He}$ (triangles), $d+\overline{d}$ (filled circles), and $\overline{d}$ (open circles); the solid (dashed) line represents the deuteron ($^{3}He$) $v_{2}$ calculated by the blast-wave model. (b) $d+\overline{d}$ and $^{3}He+\overline{^{3}He}$ $v_{2}$ as a function of $p_{T}$, both $v_{2}$ and $p_{T}$ have been scaled by $A$. Errors are statistical only. $\overline{p}$ (open diamonds) and $\Lambda+\overline{\Lambda}$ (solid stars) $v_{2}$ are shown in the plot as a comparison. Coalescence and dynamic simulation for deuteron from Ref.~\cite{coal_dyn} are also shown. (c) Low $p_{T}$ $\overline{d}$ $v_{2}/A$ (open circles) as a function of centrality fraction ($0-10\%,\ 10-20\%,\ 20-40\%,\ 40-80\%$, respectively). Errors are statistical only. $\overline{p}$ $v_{2}$ is also shown as open diamonds. Blast-wave calculations are show as solid ($\overline{d}$) and dashed lines ($\overline{p}$).The 2 subpanels are for different $p_{T}$ ranges.}
\label{flow_figure}
%\end{center}
\end{figure}

Figure~\ref{flow_figure} panel (a) shows $v_{2}$ as a function of $p_{T}$ for $d+\overline{d}$, $^{3}He+\overline{^{3}He}$ and $\overline{d}$ in minimum-bias collisions. The results with both
$v_{2}$ and $p_{T}$ scaled by $A$ are shown in panel (b). As mentioned in the previous subsection, the $\pi$, $K$, $p$ spectra and $v_{2}$ are all described by a single set of blast-wave parameters and then the spectra and $v_{2}$ of $d$($\overline{d}$), $^{3}He$($\overline{^{3}He}$) are calculated. The blast-wave results for the deuteron ($^{3}He$) $v_{2}$ are shown as the solid (dashed) line. As a comparison, the $\overline{p}$ and $\Lambda+\overline{\Lambda}$ $v_{2}$ \cite{pbar_flow} are superimposed on the plot.  Results for deuteron $v_{2}$ from coalescence and dynamic models~\cite{coal_dyn} are also shown. The data suggest that the $d+\overline{d}$ and baryon $v_{2}$ seem to follow the $A$ scaling within errors, indicating that the $d+\overline{d}$ are formed through the coalescence of $p(\overline{p})$ and $n(\overline{n})$ just before thermal freeze-out.  However, the scaled $^{3}He+\overline{^{3}He}$ $v_{2}$ appears to deviate a bit more from the blast-wave calculated $v_{2}$. To quantify the degree of agreement with the scaling, we performed a $\chi^2$ analysis by comparing the nucleus $v_2$ to the curves which describe the baryon
data. The $\chi^2$ per degree of freedom for deuteron $v_2$ is 3.1/2 while that of $\overline{^{3}He}$ is 4.1/2. There appears to be a slightly larger deviation for heavier elements.

The anti-deuteron $v_{2}/A$ as a function of centrality fraction is shown in Figure~\ref{flow_figure} panel (c). The upper and lower sub-panels represent results for two different regions of $p_{T}$. The $\overline{d}$ is observed to have a negative $v_{2}$ in central and mid-central collisions in the transverse momentum range of $0.2<p_{T}<0.7$ GeV/c. This negative $v_{2}$ is consistent with a large radial flow, as the blast-wave calculations show. At the same $p_{T}/A$ where the $\overline{d}$ is negative, the $\overline{p}$ $v_{2}$ is consistent both with zero and with the $\overline{d}$ $v_{2}$, due to the large uncertainties. The blast-wave parameters published in Ref.~\cite{pbar_flow} are used to calculate $\overline{p}$ and $\overline{d}$ flow and the calculated results are shown in the figure. The $p_T$ ranges of the $\overline{p}$ data points are selected to match approximately the same $p_T/A$ center of the $\overline{d}$ points used here. Though the blast-wave model predicts the generic feature of negative $v_{2}$, quantitative agreement between data and model throughout the entire centrality and $p_{T}$ range is lacking.

\subsection{Baryon density at $\mu_B=0$}

As discussed in the introductory section, the $d/p$ ratio is proportional to baryon density. In the collider configuration, the production of low-$p_T$ deuterons is often overwhelmed by background
deuterons from the interaction of energetic hadrons (pions and protons) with detector material close to the beam (beam pipe, etc.). When the net baryon density is close to zero, anti-deuterons can be used as a measure of deuteron production. Reference~\cite{xuliu} shows that the baryon density in $\gamma$p, pp, pA and AA collisions follows a universal distribution as a function of beam energy and can be described by statistical processes. At zero baryon chemical potential ($\mu_B=0$), the $d/p$ ratio and the $\bar{d}/\bar{p}$ ratio are identical and the measurements from all systems considered are consistent with each other. For the 5 data points, which are closest to the $\mu_B=0$ condition (Au+Au collisions at $\sqrt{s_{_{NN}}}=200$ GeV (1 point from STAR and the other from PHENIX), $e^+e^-\rightarrow ggg$ at $\sqrt{s}=10$ GeV, $\gamma p$ at $\sqrt{s}=200$ GeV, and $\bar{p}+p$ at $\sqrt{s}=1.8$ TeV), the average value is $\bar{d}/\bar{p}=(7.6\pm0.8)\times10^{-4}$. We note that the $D/H$ value of $2.8\pm0.2)\times10^{-5}$ obtained from Big Bang nucleosynthesis in the evolution of the Universe~\cite{BBN1,PDGBBN} is about 4\% of what is obtained in higher energy processes.

\section{Summary}
Using the particle identification capabilities of the STAR TPC and TOF detectors, we have measured the $d(\overline{d})$ and $^{3}He(\overline{^{3}He})$ $p_{T}$ spectra. The extracted coalescence
parameters $B_{2}$ and $\sqrt{B_{3}}$ have similar values. They have smaller values for more central collisions, which is consistent with an increasing source size with an increase in collision
centrality. The nuclei coalescence parameters are proportional to the inverse of the freeze-out volume estimated using the pion HBT radii in all centrality classes.  The spectra of nuclei with $A=2$ and 3 are in general described by the blast-wave model, which was used to describe the spectrum, elliptic flow, and HBT results.  However, the model overpredicts the radial flow, which implies that the coalescence process is different from the simple mass effect as assumed in the blast-wave model.

We have also measured the $v_{2}$ of $d(\overline{d})$ and $^{3}He(\overline{^{3}He})$. The $v_{2}$ values of $d(\overline{d})$ when scaled by atomic mass number $A$, follows the baryon $v_{2}$, thereby providing evidence of $d(\overline{d})$ formation through final-state coalescence of nucleons. We also observed the $v_{2}$ for $\overline{d}$ to be negative at low $p_{T}$ in the mid-central collisions. Comparison with blast-wave calculations shows this is consistent with a large radial flow in Au+Au collisions at a small impact parameter.

\section{Acknowledgments}
We thank the RHIC Operations Group and RCF at BNL, and the NERSC Center at LBNL and the resources provided by the Open Science Grid consortium for their support. This work was supported in part by the Offices of NP and HEP within the U.S. DOE Office of Science, the U.S. NSF, the Sloan Foundation, the DFG cluster of excellence `Origin and Structure of the Universe', CNRS/IN2P3, RA, RPL, and EMN of France, STFC and EPSRC of the United Kingdom, FAPESP of Brazil, the Russian Ministry of Sci. and Tech., the NNSFC, CAS, MoST, and MoE of China, IRP and GA of the Czech Republic, FOM of the Netherlands, DAE, DST, and CSIR of the Government of India, the Polish State Committee for Scientific Research,  and the Korea Sci. \& Eng. Foundation.

\end{document}